\documentclass[reprint, pra,aps,showpacs,letterpaper,superscriptaddress]{revtex4-1}
\usepackage{times}
\usepackage[utf8]{inputenc}
\usepackage[english]{babel}
\usepackage{bm}
\usepackage{graphicx}
\usepackage{amsmath}
\usepackage{amssymb}
\usepackage{amsfonts}
\usepackage{braket}
\usepackage{xcolor}

\begin{document}

\title{Photoelectron spectra after multiphoton ionization 
of Li atoms in the one-photon Rabi-flopping regime}

\author{D.~A.~Tumakov}
\email{dm.tumakov@gmail.com}
\affiliation{Department of Physics, St.~Petersburg State University, 
Universitetskaya Naberezhnaya 7/9, St.~Petersburg 199034, Russia}
\author{Dmitry~A.~Telnov}
\email{d.telnov@spbu.ru}
\affiliation{Department of Physics, St.~Petersburg State University, 
Universitetskaya Naberezhnaya 7/9, St.~Petersburg 199034, Russia}
\author{G.~Plunien} 
\affiliation{Institut f\"ur Theoretische Physik, Technische Universit\"at 
Dresden, Mommsenstrasse 13, Dresden D-01062, Germany}
\author{V.~M.~Shabaev}
\affiliation{Department of Physics, St.~Petersburg State University, 
Universitetskaya Naberezhnaya 7/9, St.~Petersburg 199034, Russia}
\begin{abstract}
We calculate two-dimensional photoelectron momentum distributions and energy spectra after 
multiphoton ionization of Li atoms subject to intense laser fields in one-photon Rabi-flopping regime. 
The time-dependent Schr\"{o}dinger equation is solved 
within the single-active-electron approximation using a model potential which 
reproduces accurately the binding energies and dipole matrix elements of the 
Li atom. Interaction with the external electromagnetic field is treated 
within the dipole approximation. We show that the Rabi 
oscillations of the population between the ground $2s$ state and the excited 
$2p$ state in the one-photon resonance regime are reflected in the energy 
spectra of emitted photoelectrons which manifest interference structures with 
minima. Transformations of the interference structures in the photoelectron 
energy spectra caused by the variation of the laser peak intensity and 
pulse duration are analyzed.
\end{abstract}

\maketitle

\section{Introduction}
The phenomenon of above-threshold ionization (ATI), which was discovered about 
40 years ago~\cite{agos1979}, and study of the resulting photoelectron angular 
distributions (PAD) attract much interest both theoretically and experimentally. 
Over the last four decades, this interest has grown significantly, which is 
related to the rapid progress of laser technology, namely, the possibility of 
generating extremely short and intense pulses~\cite{brabec2000}. To get the general picture of the 
problem the reader can refer to a number of review papers~\cite{becker2002,milosevic2006,agostini2012}. 

A lithium (Li) atom, being the simplest open-shell system, presents itself a 
unique target for laser-atom interaction investigations, drawing 
both experimental~\cite{zhu2009, schuricke2011} and 
theoretical~\cite{yip2010, armstrong2012, morishita2013, avanaki2016, jheng2017, murakami2018} 
attention. From a theoretical perspective, it is 
important that the lithium atom has a single electron outside 
a closed shell, which enables the single-active-electron (SAE) model to come 
into play for an accurate description of the electron 
dynamics~\cite{schuricke2011, morishita2013, jheng2017}. For all the laser 
pulse parameters considered in the present paper, it is well established that 
the time-dependent Schr\"{o}dinger equation within the SAE formulation is 
adequate for describing the ionization process~\cite{schuricke2011}.

Photoelectron spectra and 2D momentum distributions contain various information 
about the ionization process, and also about the atomic or molecular 
internal structure. While a typical long-pulse ATI energy 
spectrum exhibits a well-known structure of equally-spaced 
peaks~\cite{agos1979}, it can also have different subtle features: Stark-induced 
Rydberg states resonances (Freeman resonances)~\cite{freeman1987}, low energy 
structure (LES)~\cite{blaga2009}, or interference structure originating from 
interfering electrons, emitted at different times~\cite{telnov1995, lindner2005,
wickenhauser2006, arbo2006, arbo2010, arbo20102}. 
PAD can be efficiently calculated by means of methods involving 
partition of the whole coordinate space into two regions and analytical 
propagation of the wave function in the external region without including the 
interaction with the atomic core~\cite{chelkowski1998, tong2006, 
giovannini2012}, with the approach based on the transition to the 
Kramers-Henneberger reference frame~\cite{telnov2009}, or by calculating a 
time-dependent flux through a 
spherical surface placed far enough from the atomic core~\cite{tao2012}. For the 
processes considered in the present paper, however, the resulting photoelectrons 
have rather small kinetic energies (about 0.05-0.5~eV), which makes it 
unreasonable to neglect the Coulomb potential in the final 
states~\cite{chen2006, blaga2009, telnov2011}. For the calculation of the PAD we are 
providing a simulation box which is large enough to capture the dynamics of an 
ionized electron, which allows us to obtain the PAD directly projecting the 
final wave function onto the unbound states built with the scattering theory 
methods, including the interaction with the atomic core~\cite{wiehle2003, 
chen2006}.

For decades since the pioneering Rabi work~\cite{rabi1937}, population transfer 
between two electron states, induced by a resonant external electromagnetic 
field, has been a powerful tool of controlling quantum systems. In the recent 
paper~\cite{avanaki2016} the high-order-harmonic generation (HHG) of a Li atom 
was studied in one- and two-photon Rabi-flopping regimes, revealing multipeak 
oscillatory pattern emerging in HHG spectra, which corresponds directly to the 
coherent population transfer between the ground $2s$ state and the excited $2p$, 
$3s$ and $3d$ states. In the present paper, we calculate photoelectron energy 
spectra and angular distributions of a Li atom in Rabi-flopping regime. We 
show that oscillations of the population of an excited state lead to the 
emergence of a prominent interference structure in the resulting photoelectron 
spectra. 

The paper is organized as follows. In Sec.~\ref{sec:theory} we describe in  
detail the theoretical and computational methods 
applied to the present problem. The results of our calculations and all 
necessary theoretical analyses are presented in Sec.~\ref{sec:results}. 
Sec.~\ref{sec:summary} summarizes the results. Atomic units are used throughout 
the paper ($\hbar = m = e = 1$), unless specified otherwise.

\section{Theoretical and computational methods}
\label{sec:theory}
\subsection{Electronic structure of Li atom}
For the description of unperturbed electronic states of the 
lithium atom within SAE, we make use of the Klapisch model 
potential~\cite{klapisch1971}:
\begin{equation}\label{eq:Klapisch}
V_{\mathrm{K}}(r) = - \frac{1}{r} \left ( 1 + \left ( Z - 1 \right )  
\mathrm{e}^{-\alpha r} + C r \mathrm{e}^{- \beta r} \right ),
\end{equation}
where $Z$ is the nucleus charge (for lithium, $Z = \mathrm{3}$). Other parameters are 
taken from Ref.~\cite{magnier1999}:
\begin{equation}
\alpha =  \text{7.90875}, \; \beta = \text{3.90006}, \; C = \text{10.321}.
\end{equation}
The eigenvalues and eigenfunctions of the unperturbed 
one-electron Hamiltonian are obtained by solving the time-independent 
Schr\"odinger equation in spherical coordinates. Since the atomic core 
potential is spherically symmetric, the eigenfunctions take a form with 
separated radial and angular coordinates:
\begin{equation}\label{eq:stacStates}
\psi_{nlm} (\bm{r}) = R_{nl}(r)Y_{lm} (\theta, \varphi).
\end{equation}
Here $Y_{lm} (\theta, \varphi)$ are the spherical harmonics 
with $l$ and $m$ being the angular momentum and its projection. The radial 
eigenfunctions $R_{nl}(r)$ are enumerated with the index $n$ for each $l$. They 
satisfy the following equations:
\begin{equation}\label{eq:stacSchrod}
H_0^l R_{nl}(r) = \epsilon_{nl} R_{nl}(r),
\end{equation}
\begin{equation}
H_0^l = - \frac{1}{2}\left[\frac{\partial^2}{\partial r^2} + 
\frac{1}{r}\frac{\partial}{\partial r} \right]+ \frac{l (l + 1)}{2 r^2} 
+ V_{\mathrm{K}}(r).
\end{equation}
The equations are solved with the 
generalized pseudospectral (GPS) method (for the details, see, 
for example, \cite{yao1993,telnov1999,telnov2013}). In this method, the 
Hamiltonian operator and wave functions are discretized on a nonuniform radial 
grid, and the resulting matrix eigenvalue problem for each value of $l$ is 
solved efficiently with the standard linear algebra routines.

For an accurate description of the resonant processes involving the initial $2s$ state and higher-lying 
excited states, the model must
provide accurate excitation energies and dipole transition matrix elements. 
To make sure this is the case for the Klapisch potential, we calculate these quantities for 
several excited energy levels of the lithium atom. The 
excitation energies are listed in Table~\ref{table:excitation}. As one can see, 
they agree very well with the experimental data from Refs.~\cite{anwar2005, 
radziemski1995}.

\begin{table}
\caption{Ionization and excitation energies of Li (in 
atomic units) calculated with the Klapisch model 
potential (Eq.~\eqref{eq:Klapisch}) in comparison with the experimental data from 
Refs.~\cite{anwar2005,radziemski1995}.}
\label{table:excitation}
\centering
\begin{ruledtabular}
\begin{tabular}{lcc} 
 Transition & Model potential & Experiment \\ %[0.5ex] 
 \hline
 $2s \rightarrow \text{continuum}$ & 0.198\phantom{1} & 0.198\phantom{1} \\ 
 $2s \rightarrow 2p$ & 0.0679 & 0.0679 \\
 $2s \rightarrow 3s$ & 0.1238 & 0.1240 \\
 $2s \rightarrow 3p$ & 0.1408 & 0.1409 \\
 $2s \rightarrow 3d$ & 0.1425 & 0.1425 \\ [1ex] 
\end{tabular}
\end{ruledtabular}
\end{table}

\begin{table}
\caption{Transition dipole matrix elements 
$\braket{n'l'|z|nl}$ of Li (in 
atomic units) calculated with the Klapisch model 
potential (Eq.~\eqref{eq:Klapisch}) in comparison with the matrix elements calculated by the precision linearized 
coupled-cluster method~\cite{safronova2012}.}
\label{table:transme}
\centering
\begin{ruledtabular}
 \begin{tabular}{lcc} 
 Transition & Model potential & Coupled-cluster method~\cite{safronova2012} \\ %[0.5ex] 
 \hline
 $2s \rightarrow 2p$ & 2.35\phantom{1} & 2.35\phantom{1} \\
 $2s \rightarrow 3p$ & 0.129 & 0.129 \\
 $2p \rightarrow 3s$ & 1.72\phantom{1} & 1.72\phantom{1} \\
 $2p \rightarrow 3d$ & 2.26\phantom{1} & 2.27\phantom{1} \\ [1ex] 
\end{tabular}
\end{ruledtabular}
\end{table}

The same is also true for the corresponding transition dipole matrix elements presented in Table~\ref{table:transme}. 
They show excellent agreement with the matrix elements obtained by the precision linearized 
coupled-cluster method~\cite{safronova2012}.

\subsection{Time propagation of the wave function}
To obtain the time-dependent wave function of the active 
electron $\Psi(\bm{r}, t)$ in the laser field, we solve the time-dependent 
Schr\"{o}dinger equation (TDSE):
\begin{equation}\label{eq:tdse}
i \frac{\partial \Psi(\bm{r}, t)}{\partial t} = \left[ H_{0} + V(\bm{r}, t) 
\right] \Psi(\bm{r}, t),
\end{equation}
\begin{equation}\label{eq:H0}
 H_{0} =  -\frac{1}{2} \nabla^2 + V_{\mathrm{K}}(r),
\end{equation}
for the initial $2s$ electronic state:
\begin{equation}\label{eq:init}
\Psi(\bm{r}, 0) = \psi_{2s} (\bm{r}).
\end{equation}
The interaction with the external electromagnetic field is treated within the 
dipole approximation, which is well justified for the laser field intensities 
and wavelength used in the present calculations (see, for example, 
Ref.~\cite{ludwig2014}). In the length gauge, 
the interaction potential $V(\bm{r}, t)$ takes the form:
\begin{equation}
V(\bm{r}, t) = \bm{F}(t) \cdot \bm{r},
\end{equation}
where $\bm{F}(t)$ is the electric field strength. We assume 
linear polarization of the laser field along the $z$ axis:
\begin{equation}\label{eq:field}
\bm{F}(t) = \bm{e}_z F_0 f(t) \sin \omega t,
\end{equation}
where $\omega$ and $F_0$ are the carrier frequency and peak field strength, 
respectively, while $f(t)$ is a slowly varying pulse envelope. We make use of 
the trapezoidal pulse shape with smooth edges to reduce the effects of varying 
intensity on the leading and trailing edges of the laser pulse:
\begin{equation}\label{eq:modelenv}
f(t) = 
\begin{cases}
\sin^{2}\left(\dfrac{\pi t}{2\Delta T} \right)  & \mathrm{if}\quad 0 
\leq t \leq \Delta T, \\
1 & \mathrm{if}\quad \Delta T \leq t \leq T - \Delta T, \\
\sin^{2}\left(\dfrac{\pi (t - T)}{2\Delta T} 
\right) & \mathrm{if}\quad T - \Delta T \leq t \leq T, \\
0 & \mathrm{if}\quad t<0\quad\mathrm{or}\quad t>T,
\end{cases}
\end{equation}
where $\Delta T = 0.1 T$ is the switching duration, and $T$ is the pulse 
duration. For a pulse containing $N$ optical cycles of the 
frequency $\omega$, it is defined as
\begin{equation}
T = \frac{2 \pi N}{\omega}.
\end{equation}
In the present work we set the carrier wavelength to 671~nm (photon energy 
0.0679~a.u.) which corresponds to a resonant one-photon transition between $2s$ 
and $2p$ states, and the peak intensities vary in the range from $10^{10}$ to 
$10^{12}$~W/cm$^2$. The laser pulse contains 20~optical cycles (duration 
is about 44 fs) unless specified otherwise.

To solve Eq.~\eqref{eq:tdse} numerically, we apply the 
time-dependent general pseudospectral (TDGPS) method~\cite{tong1997} (for the 
details of the method, see also \cite{telnov2013, murakami2017}). Here we 
briefly outline our computational procedure. For the external field 
\eqref{eq:field} linearly polarized along the $z$ axis, the projection of the 
electron angular momentum on this axis is conserved. Then the time-dependent 
wave functions can be expanded on the basis of spherical harmonics with $m=0$: 
\begin{equation}\label{eq:wfexp}
\Psi(\bm{r}, t) = \sum_{l = 0}^{l_{\mathrm{max}}} g_l(r, t) Y_{l 0} 
(\theta, \varphi).
\end{equation}
Here $l_{\mathrm{max}}$ is the maximum angular momentum used in the 
calculations. For the initial state~\eqref{eq:init}, one has 
\begin{equation}
g_l(r, 0) = R_{20}(r) \delta_{l0}.
\end{equation}
The time propagation method is based on the split-operator technique in the 
energy representation~\cite{tong1997}. The short-time propagator is defined by 
the following expression:
\begin{equation}\label{eq:split}
\begin{split}
\Psi(\bm{r}, t +\Delta t) &=  \exp\left[-i\frac{\Delta 
t}{2}H_{0}\right]\\
&\times\exp\left[-i \Delta t V\left(t + \frac{\Delta t}{2}\right)\right]\\
&\times \exp\left[-i\frac{\Delta t}{2}H_{0}\right] \Psi(\bm{r}, t),
\end{split}
\end{equation}
where $\Delta t$ is the time step and
\begin{equation}\label{eq:propunp}
\begin{split}
\exp\left[-i\frac{\Delta t}{2}H_{0}\right] = \sum_{l=0}^{l_{\mathrm{max}}}|Y_{l 0}\rangle \exp\left[-i\frac{\Delta t}{2}H_0^l \right] \langle Y_{l 0}|.
\end{split}
\end{equation}
Eq.~\eqref{eq:split} is applied recursively 
starting at $t=0$ until the final wave function is obtained at $t=T$. As 
one can see from Eq.~\eqref{eq:propunp}, the unperturbed propagator is actually 
reduced to the radial propagators corresponding to the individual angular momenta 
$l$, thus the angular momentum representation of the wave function 
\eqref{eq:wfexp} perfectly suits this propagation method. The 
unperturbed propagators $\exp\left[-i(\Delta t/2)H^l_{0}\right]$ are
time-independent and need to be calculated only once before the propagation 
procedure starts. For this purpose, we use the spectral expansion:
\begin{equation}
\exp\left[-i\frac{\Delta t}{2}H_0^l\right] = \sum_n \exp\left(-i\epsilon_{nl}\frac{\Delta t}{2} \right)| R_{nl} \rangle\langle R_{nl} |.
\end{equation}
Using this expansion, we can also control the contributions of extremely high 
energy states (large $n$), which are irrelevant for the physical processes 
under consideration, improving the numerical stability of the computations. 
The matrix dimensions of 
the radial propagators could be much smaller than that of the total 
propagator depending on the largest angular momentum $l_{\mathrm{max}}$ used. On 
the contrary, the external field part of the total propagator, $\exp\left[-i 
\Delta t V\left(t + \Delta t/2\right)\right]$, is best calculated in the 
coordinate ($r$ and $\theta$) representation where it appears as a 
multiplication operator. As any multiplication operator is diagonal in the GPS 
method, its calculation is not time-consuming, even though it is 
time-dependent and must be calculated at each time step. For the present 
numerical scheme to work, the wave function has to be transformed forth and back 
between the angular momenta $l$ and coordinate $\theta$ representations at each 
time step. Of course, such transformations take additional computer time but it 
is well compensated by the speedup due to the optimal propagator representation.

For the highest laser peak intensity 5.5 $\times 
10^{11}$~W/cm$^2$ used in our calculations, the numerical parameters are as 
follows: the largest angular momentum is $l_{\mathrm{max}}=$ 50, the simulation 
box size is $R_{\mathrm{max}} = $ 400~a.u., the number of radial grid points is 
311, and the number of time steps per optical cycle is 5000. For lower 
intensities, the parameters values may be reduced. The GPS 
discretization assumes zero boundary conditions for the wave function at 
$r=R_{\mathrm{max}}$. We do not use any special absorber at large distances to 
prevent spurious reflections from the box boundary. In this paper, we study the 
photoelectron distributions within the first ATI peak only, and the 
components of the wave packet corresponding to the first ATI peak (with the 
energies less than $0.023$~a.u.) do not reach the box boundary by the end of the 
laser pulse. The components with higher energies may be reflected but they are 
very small (the second ATI peak is 2 to 3 orders of magnitude weaker than the 
first one) and do not affect the dynamics of the process.
Convergence of the results has been checked with respect to variation of the 
numerical parameters. Some calculations have also been performed using the 
velocity gauge of the interaction with the laser field and confirmed reliability 
of the corresponding length gauge results.

\subsection{ATI electron spectra}
At the end of the laser pulse, the transition amplitudes to 
various electronic states can be obtained by projecting the final wave function 
onto the corresponding eigenfunctions of the unperturbed Hamiltonian. We should 
note here that the initial wave function represents an excited ($2s$) state of 
the model SAE Hamiltonian, so the transitions to the ground $1s$ state are also 
possible. This is an obvious deficiency of the SAE model, since in the real Li 
atom the $1s$ state is occupied by two electrons with 
opposite spins, and 
transitions to this state are not permitted by the Pauli principle. For the 
laser field frequency and intensities used in the present calculations, 
however, the spurious transitions to the $1s$ state after the laser pulse are 
negligibly small. For relatively weak fields, the 
probabilities of multiphoton transitions depend strongly on the number of 
photons involved in the process. The probability of the 
process drops rapidly as the number of photons increases. In our case, 3 
photons are required to ionize the $2s$ state while 27 photons must 
be emitted when a transition from the $2s$ state to the $1s$ state (unoccupied 
in our one-electron model) occurs. One can expect that the probability of this 
transition would be very small compared to the ionization probability of the 
$2s$ state. Our results confirm that the population of the $1s$ state 
remains negligibly small after the laser pulse for all the intensities used in 
the calculations. For the same reason, excitation and ionization probabilities 
of the $1s$ electronic shell in the real Li atom are also extremely small, so we 
can draw a conclusion that the SAE model properly describes the physical 
processes under consideration. The situation may change if different laser 
field parameters are used. Transitions between the $1s$ and other states may 
become noticeable for the field with the higher intensity and/or higher 
frequency, and this would indicate the breakdown of the present SAE model. Since 
fully \textit{ab initio} three-electron calculations are not feasible at this 
time, one may think about using approximate multielectron approaches, such as a 
model of independent electrons moving in a mean field with the wave function 
described by a Slater determinant or the time-dependent density functional 
theory with some approximate exchange-correlation functional. However, 
multielectron approaches are beyond the scope of this paper because they are not 
really required by the physics for the laser field parameters used in the 
present calculations.

The photoelectron angular and energy (or momentum) 
distribution (PAD) can be calculated at the end of the laser pulse by projecting 
the wave function $\Psi(\bm{r}, T)$ onto the continuum eigenfunctions 
$\Psi^{-}(\bm{k}, \bm{r})$ of the unperturbed Hamiltonian, corresponding to 
the energy $E = k^2 / 2$ and asymptotic momentum direction~$\hat{\bm{k}}$. 
The functions $\Psi^{-}(\bm{k}, \bm{r})$ can be represented by the partial 
wave expansion \cite{newton}:
\begin{equation}\label{eq:final}
\Psi^{-}(\bm{k},\bm{r}) = \sqrt{\frac{2}{\pi}} \sum_{l = 0}^{l_{\mathrm{max}}} 
\sum_{m = - l}^l i^l e^{- i \delta_l} \psi_l(k, r) Y^*_{lm} (\hat{\bm{r}}) 
Y_{lm} (\hat{\bm{k}}),
\end{equation}
where $\delta_l$ are the scattering phase shifts. The partial waves $\psi_l 
(k, r)$ satisfy the following equation:
\begin{equation}\label{eq:shifts}
H_0^l \psi_l(k, r) = \frac{k^{2}}{2}\psi_l(k, r)
\end{equation}
and should be normalized according to the asymptotic form at 
$r\rightarrow\infty$:
\begin{equation}
\psi_l(k, r) \approx \frac{1}{kr} \sin \left(kr - \frac{\pi l}{2} +\frac{1}{k} 
\ln(2kr) + \delta_l \right).
\end{equation}
Eq.~\eqref{eq:shifts} is solved by the finite-difference Numerov 
method (using 
a power series expansion of the function $\psi_l (k, r)$ in the vicinity of 
$r = 0$), providing both the partial waves and phase shifts. Then the full wave 
function $\Psi^{-}(\bm{k},\bm{r})$ is constructed according to 
Eq.~\eqref{eq:final}.

The differential ionization probability for the electrons 
emitted with the momentum $\bm{k}$ into the unit energy and solid angle 
intervals is calculated as
\begin{equation}\label{eq:pad}
\frac{dP(\bm{k})}{dE d \Omega} = k |\braket{\Psi^{-}(\bm{k}, \bm{r}) | 
\Psi(\bm{r}, T)}|^2.
\end{equation}
Photoelectron energy spectrum can be obtained by integration of
PAD~\eqref{eq:pad} over the angles:
\begin{equation}\label{eq:pes}
\frac{dP(E)}{dE} = \int \frac{dP(\bm{k})}{dE d \Omega} d \Omega.
\end{equation}
Then the total ionization probability can be calculated performing additional 
integration of the spectrum~\eqref{eq:pes} over the emitted electron energy:
\begin{equation}\label{eq:full2}
P = \int_{0}^{\infty} \frac{dP(E)}{dE} dE.
\end{equation}
The same quantity can be obtained by projecting the wave function $\Psi(\bm{r}, 
T)$ onto the eigenstates of the unperturbed discretized Hamiltonian with 
positive energies:
\begin{equation}\label{eq:full1}
P = \sum_{\epsilon_{nl} > 0} |\braket{\psi_{nl}(\bm{r}) | \Psi(\bm{r},T)}|^2.
\end{equation}
Strictly speaking, the results returned by Eqs.~\eqref{eq:full2} and 
\eqref{eq:full1} are not identical, since the wave functions \eqref{eq:final} 
are not the eigenstates of the \textit{discretized} unperturbed Hamiltonian. 
However, when the numbers of radial grid points and angular momenta are large 
enough to ensure convergence of the calculations, these results must be close 
to each other. In all our calculations the difference between the ionization 
probabilities obtained with Eq.~\eqref{eq:full1} and Eq.~\eqref{eq:full2} is 
less than 0.5\%.

\section{Results and discussion}\label{sec:results}
A non-resonant ionization process in the presence of an intense laser field is 
usually characterized in terms of the tunneling ionization (TI) and multiphoton 
ionization (MPI) models. The separation of these two regimes is related to the 
value of Keldysh parameter~\cite{keldysh1965} $\gamma = \sqrt{I_p / 2 U_p}$, 
with $U_p = F_0^2 / 4 \omega^2$ being the ponderomotive energy for 
linearly-polarized field, and $I_p$ being the electron ionization potential. A 
slowly varying strong field corresponds to $\gamma \ll 1$ and the TI model, 
$\gamma \gg 1$ corresponds to the MPI regime. In our present calculations, the 
Keldysh parameter varies from 8 to 80, which restricts us to the latter case. 
Photoelectron spectrum within this model contains (in the weak-field regime) 
equally-spaced peaks separated by a photon energy value (sharp peaks for a 
monochromatic external field)~\cite{agos1979}. Positions of these peaks can be 
roughly estimated as $E_n = - I_p + n \omega - U_p$, so with the increasing 
intensity the peaks are supposed to shift to the lower energies region.

In the present study, we set the laser wavelength to 
671~nm, which matches the one-photon resonance between $2s$ and $2p$ states 
of the Li atom described with the Klapisch model 
potential (the experimental value is also 671~nm). For a description of the 
Rabi oscillations between the two states, we introduce the 
Rabi frequency and the pulse area. The Rabi frequency $\Omega$ is defined as 
$\Omega = F_0 d$, where $F_0$ is an electric field strength, and $d$ is the 
transition dipole matrix element between the resonant atomic states. 
The pulse area $\Theta$ is defined as a product of the Rabi 
frequency and the full width at the half maximum (FWHM) of the laser pulse 
$\tau$: $\Theta = \Omega \tau$. For the  
trapezoidal envelope~\eqref{eq:modelenv}, $\tau = T - 
\Delta T$. When the pulse area reaches the value of $\pi$ (the $\pi$ pulse), 
the population of the initially occupied $2s$ state is
completely transferred to the $2p$ state.

\begin{figure}
\includegraphics[width=1\linewidth]{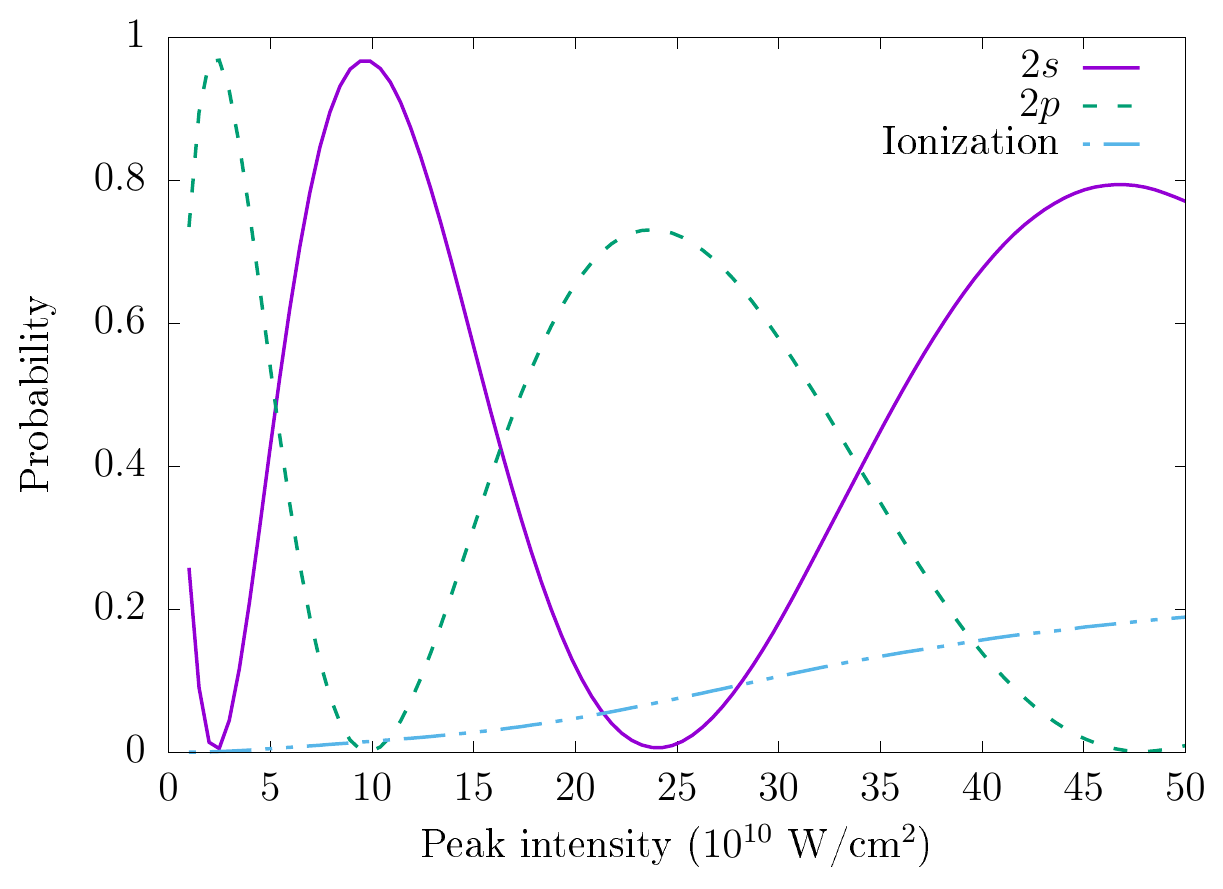}
\caption{Ionization probability and final population of $2s$ and $2p$ states of  
a Li atom exposed to linearly-polarized laser pulse. Carrier wavelength is 671~nm, 
which corresponds to resonance between $2s$ and $2p$ states. The pulse contains 
20 optical cycles, pulse envelope is given by Eq.~\eqref{eq:modelenv}.}
\label{fig:iDep}
\end{figure}

We present in Fig.~\ref{fig:iDep} the intensity dependence of the final $2s$ and $2p$ 
populations as well as the total ionization probability. Rabi oscillations between 
the $2s$ and $2p$ states are clearly 
visible on the picture. Within the intensity range we used for all our 
calculations, the total ionization probability does not exceed 0.3. Such a 
choice of the intensity range guarantees no substantial 
ionization on the leading edge of the pulse. At higher 
intensities, other excited states, namely $3s$ and $3d$ states, begin to play a 
significant role in the ionization process. With larger ac Stark 
shifts~\cite{delone1999}, resonant transitions to these states via two-photon 
absorption by the $2s$ electron come into play, making the ionization process 
more complicated.
 
\begin{figure}
\includegraphics[width=1\linewidth]{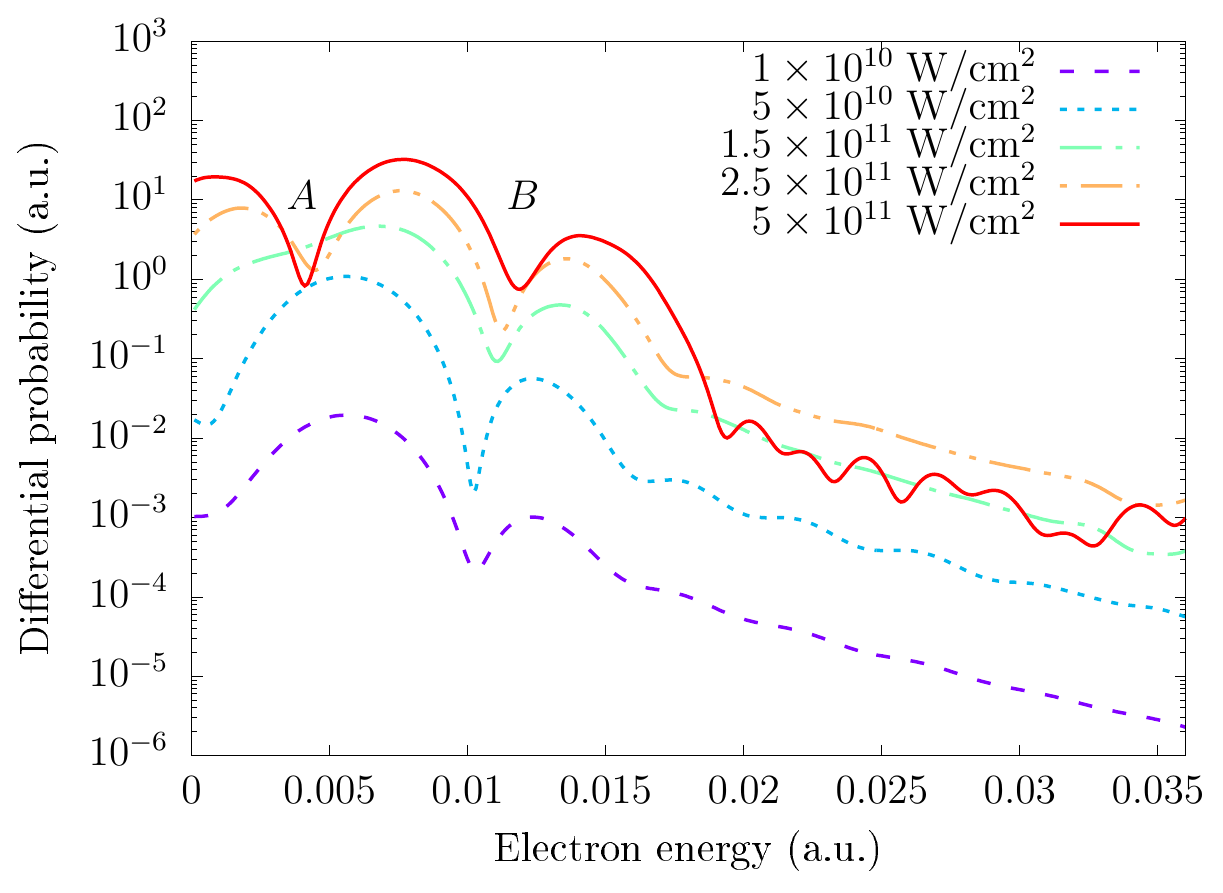}
\caption{Photoelectron spectra of a Li atom exposed to a linearly-polarized laser pulse for different laser peak intensities. The laser wavelength is 671~nm, the pulse contains 20 optical cycles. Pulse envelope is given by Eq.~\eqref{eq:modelenv}.}
\label{fig:spectra_h}
\end{figure}

Photoelectron energy spectra calculated for several laser 
peak intensities are shown in Fig.~\ref{fig:spectra_h}. In 
the multiphoton ionization regime under consideration, the first ATI peak 
(corresponding to absorption of 3 photons) is by far dominant in the 
spectrum; in what follows, we will focus on the energy distribution within 
this peak. Here one can see a clear interference pattern that builds up in the
spectrum with the increase of intensity (hence the Rabi pulse area $\Theta$) 
with two stable minima emerging near the energies of 0.01 and 0.005~a.u., 
labeled on the picture as $A$ and $B$.

\begin{figure}
\includegraphics[width=1\linewidth]{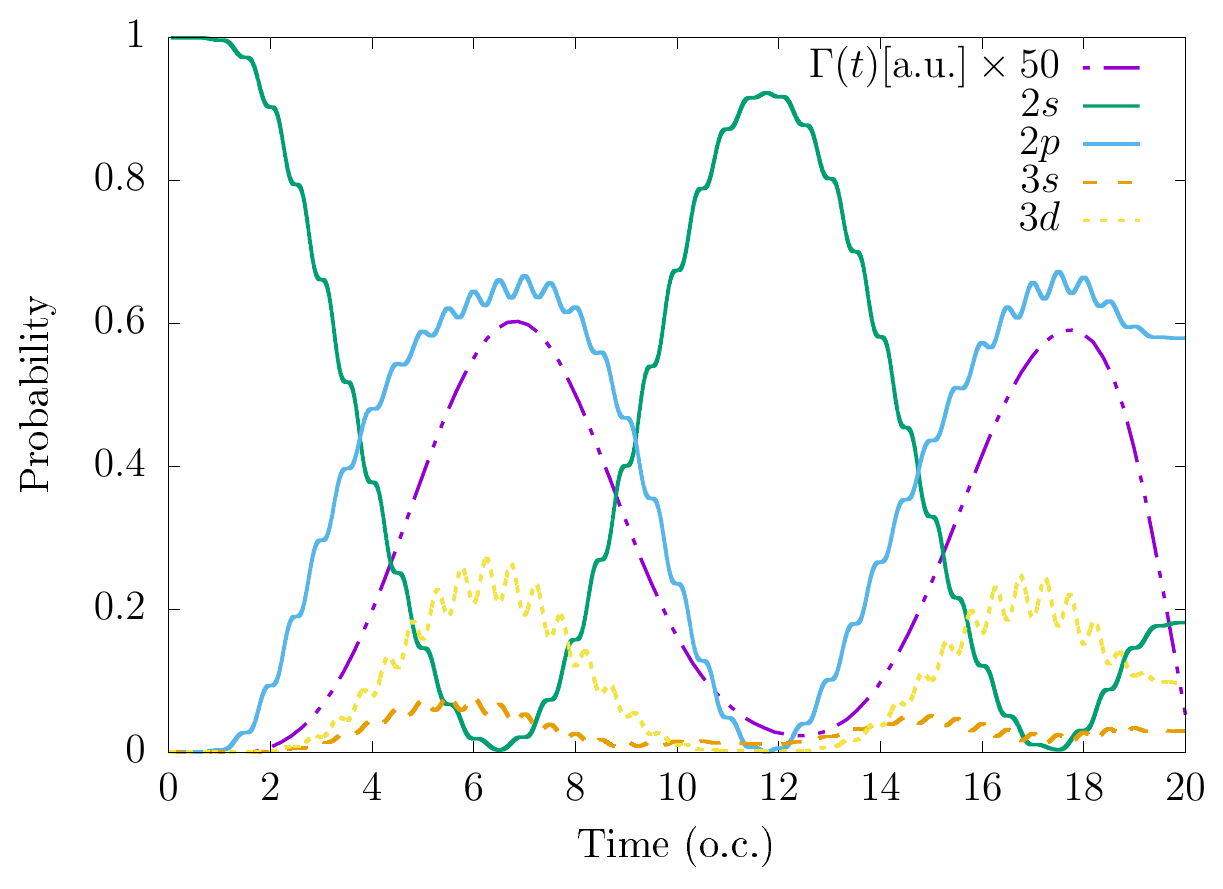}
\caption{Scaled ionization rate $\Gamma(t)$ and the time-dependent population 
of the unperturbed states of a Li atom 
exposed to a linearly-polarized laser pulse. Laser field peak intensity is 3 
$\times$ 10$^{11}$ W/cm$^2$, laser wavelength is 671~nm, which corresponds to 
resonance between $2s$ and $2p$ states. The pulse contains 20 optical cycles (o.c.), 
pulse envelope is given by Eq.~\eqref{eq:modelenv}.}
\label{fig:occup1}
\end{figure}

To reveal the origin of these minima and obtain quantitative 
estimates of their positions, one can look at the time-dependent $2s$ and $2p$ 
populations. According to the definition of $\Theta$, the  
final populations of the $2s$ and $2p$ states oscillate as 
the peak value of the electric field and/or the 
pulse duration increase. In the multiphoton ionization 
regime ($\gamma \gg 1$), the ionization probability has a sharp dependence on 
the number of absorbed photons. Ionization from the $2p$ state requires 
absorption of $2$ photons, and we expect this ionization channel to be dominant 
when the $2p$ state is populated, because $3$ photons are still required to 
ionize the Li atom directly from the ground $2s$ state. Let us define the 
time-dependent populations of the unperturbed states $\psi_{nl}$:
\begin{equation}\label{eq:tdpop}
 P_{nl}(t) = |\braket{\psi_{nl}(\bm{r}) | \Psi(\bm{r}, t)}|^2,
\end{equation}
and ionization rate as a time derivative of the instantaneous 
unbound-states population:
\begin{equation}\label{eq:tdrate}
\Gamma (t) =  \frac{d}{dt} \sum_{\epsilon_{nl} > 0} 
 P_{nl}(t).
\end{equation}
We note that the quantities defined by Eqs.~\eqref{eq:tdpop} 
and~\eqref{eq:tdrate} are not observable when the external field is still on; 
however, they may be used to illustrate the dynamics of the excitation and 
ionization processes. For the laser peak intensities such as $\Theta > 3 
\pi$ (for the trapezoidal pulse envelope~\eqref{eq:modelenv}, it corresponds to 
$I > 2.5 \times 10^{11}$~W/cm$^2$), the rate $\Gamma(t)$ exhibits two distinct 
maxima corresponding to the peaks of the time-dependent $2p$-state population.
Fig.~\ref{fig:occup1} shows the time-dependent probabilities of some unperturbed 
states for the laser peak intensity $I = 3 \times 10^{11}$~W/cm$^2$. We also 
show the averaged scaled ionization rate $\Gamma(t)$. The maxima of 
the ionization rate are certainly correlated to the maxima of the $2p$-state 
population. The second (right) maximum of the ionization rate is also 
influenced by the right edge of the pulse envelope where the instantaneous 
intensity as well as ionization rate drop rapidly. The dominant ionization 
channel is thus controlled by the population transfer between the resonant 
$2s$ and $2p$ states, which ``opens'' and ``closes'' this channel with the Rabi 
frequency $\Omega$, even when the laser field intensity remains constant.

The pattern in the ATI energy spectra 
(Fig.~\ref{fig:spectra_h}) can be qualitatively understood based on the lowest 
order of the time-dependent degenerate state perturbation theory. First, the 
zeroth order approximation for the wave function is obtained 
non-perturbatively, when the two-level system of strongly coupled resonant 
($2s$ and $2p$) states is solved using the rotating wave approximation. The 
$2s$ and $2p$ populations oscillate with the Rabi frequency:
\begin{equation}\label{eq:td2s2p}
 P_{2s}(t) = \cos^{2}\left(\frac{1}{2}\Omega t\right),\quad P_{2p}(t) = 
\sin^{2}\left(\frac{1}{2}\Omega t\right).
\end{equation}
The probability amplitudes of subsequent excitation and ionization are 
described by the corresponding multiphoton matrix elements of the perturbation. 
In the multiphoton ($\gamma\gg1$) regime, the ionization probability of the 
$2p$ state is much larger than that of the $2s$ state because ionization of the 
$2s$ state requires absorption one extra photon. For the range of the laser 
intensities under consideration, there are only two time moments within the 
pulse when the population of the $2p$ state reaches its maximum. According 
to Eq.~\eqref{eq:td2s2p}, they are separated by the time interval 
equal to $2 \pi / \Omega$.
In the vicinities of these time moments one can see the highest ionization 
rate in Fig.~\ref{fig:occup1}. The phase difference $\Phi_{\mathrm{R}}(E)$ 
between the 
contributions to the ionization amplitude from these time moments follows from 
the time dependence of the multiphoton ionization amplitude:
\begin{equation}\label{eq:phase}
\Phi_{\mathrm{R}}(E) = 2\pi(E - \epsilon_{2p} - n \omega)/\Omega,
\end{equation}
where $E$ is the energy of the ejected electron and $n = 2$ is the number of 
photons absorbed. Eq.~\eqref{eq:phase} predicts an interference structure in 
the electron energy spectra with the adjacent minima $A$ and $B$ separated by 
the Rabi frequency $\Omega$. We note that the interference oscillatory structure 
of ATI peaks can only be detected for laser pulses of finite duration; for 
continuous wave laser fields, the ATI energy spectrum in the resonant ionization 
case consists of two series of equally spaced narrow peaks shifted from each 
other by the Rabi frequency.

These predictions based on a simple model of Rabi 
oscillations in the two-level system can be checked against the results of our 
numerical calculations for the positions $E_A$ and $E_B$ of the minima $A$ and 
$B$ in the energy spectrum. Let us define two phase shifts, $\Delta 
\Phi_{\mathrm{R}}$ and $\Delta \Phi$:
\begin{eqnarray}\label{eq:dedtr}
\Delta\Phi_{\mathrm{R}} &=& 2 \pi (E_B - E_A) / \Omega,\\
\label{eq:dedt}
\Delta\Phi &=& (E_B - E_A) \Delta t,
\end{eqnarray}
where the time delay $\Delta t$ is obtained directly from the peak positions 
of the calculated time-dependent ionization rate $\Gamma(t)$: while 
$\Delta\Phi_{\mathrm{R}}$ makes 
use of the theoretically predicted time delay $2 \pi / \Omega$ in the 
two-level system, $\Delta\Phi$ is calculated with the numerical data.
Values of $\Delta\Phi$ and $\Delta\Phi_{\mathrm{R}}$, calculated for the 
laser intensities from $2.5\times 10^{11}$ to $5.5\times 10^{11}$~W/cm$^2$ 
are 
listed in Table~\ref{table:dedt}. As one can see, the phase difference between 
the two adjacent minima in the energy spectra is close to $2 \pi$ for both 
$\Delta\Phi$ and $\Delta\Phi_{\mathrm{R}}$, as expected from the interference 
pattern. 
The deviation from $2 \pi$ becomes larger for $\Delta\Phi_{\mathrm{R}}$ at 
higher 
intensities. This result is well understood: the two-level system approximation 
becomes less accurate for higher intensities as the other higher excited 
electronic states begin to play a more important role in the ionization process.

The interference oscillations in Fig.~\ref{fig:spectra_h} show 
up on top of the ATI peak in the energy range 0 to 0.02~a.u. where the 
ionization signal is the strongest and comes through the dominant ionization 
channel related to the resonant population transfer to the $2p$ state. As one 
can see, these oscillations disappear for the energies higher than 0.02~a.u. 
The energy range $0.02$ to $0.035$~a.u. in Fig.~\ref{fig:spectra_h} lies already 
beyond the first ATI peak, and the ionization signal here is 2 to 3 orders 
of magnitude weaker than that at the top of the peak. Various parts of the 
laser pulse in the time domain (including those where the $2p$ state is not 
significantly populated) can make comparable contributions to the ionization 
signal here. Since the ionization channel responsible for the interference 
oscillations is not dominant in this energy range, one cannot expect to see a 
clear interference pattern here.

\begin{figure}
\includegraphics[width=1\linewidth]{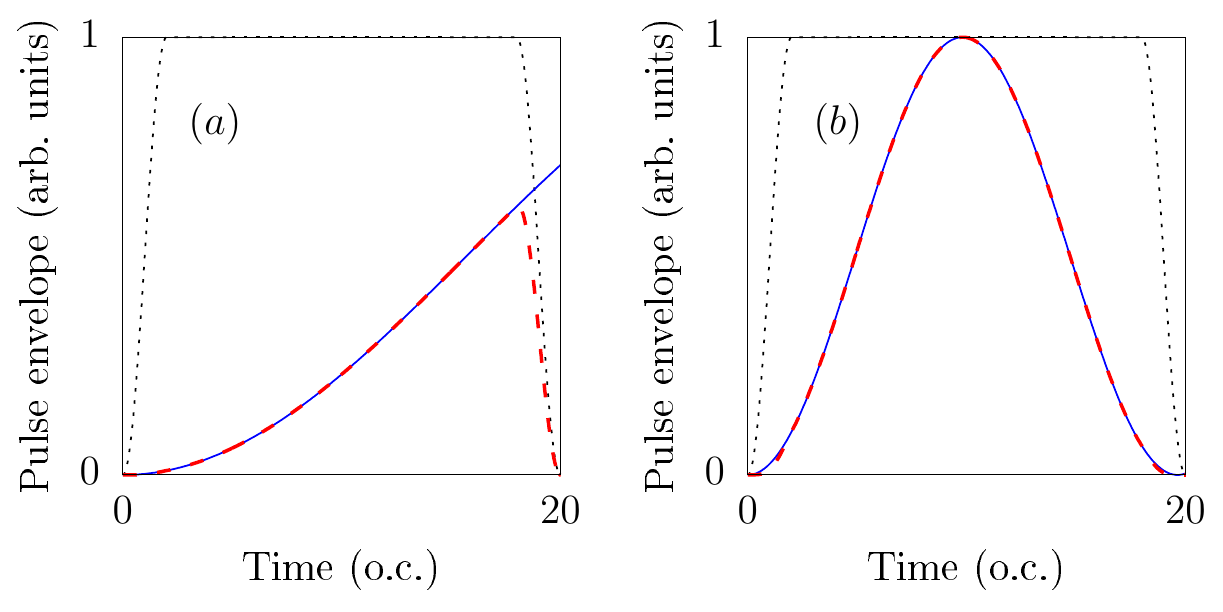}
\caption{The laser pulse envelope (Eq.~\eqref{eq:modelenv}) (dotted line), the Li $2p$ state population (solid line), and the 
schematic plot of the resulting ionization rate $\Gamma(t)$ (dashed line) for the Rabi pulse area $(a)$ $\Theta  = 0.6 \pi$ and $(b)$ 
$\Theta  = 2 \pi$. The pulse contains 20 optical cycles (o.c.).}
\label{fig:ex}
\end{figure}

\begin{figure}
\includegraphics[width=1\linewidth]{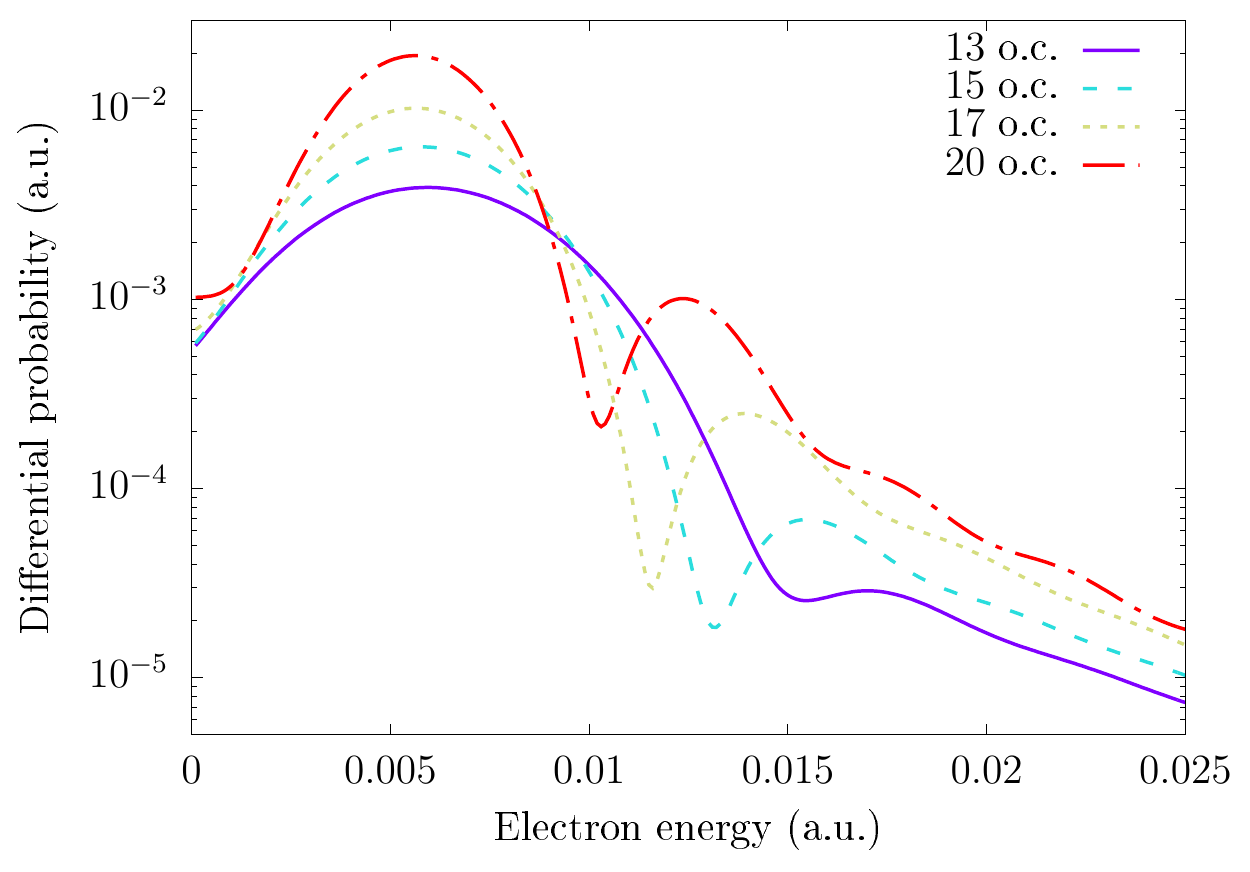}
\caption{Photoelectron spectra of a Li atom exposed to a linearly-polarized 
laser pulse for various pulse duration (13-20 optical cycles (o.c.)). 
The laser peak intensity $I =  
10^{10}$ W/cm$^2$, carrier wavelength is 671~nm, pulse envelope is given by 
Eq.~\eqref{eq:modelenv}.}
\label{fig:spectra_l}
\end{figure}

For the intensities smaller than $2.5\times 10^{11}$~W/cm$^2$, 
corresponding  to the Rabi pulse area $\Theta < 3 \pi$ for the 
pulse containing 20 optical cycles, only one peak appears in the 
ionization rate, controlled by the ionization from significantly populated 
$2p$ state (it can also be influenced by the pulse envelope edge, see Fig.~\ref{fig:ex}).
Let us introduce the width of the peak $\delta t$ and consider the 
interference of the electrons ionized during this time interval. 
Assuming the ionization rate to be equal to some mean value 
$\overline{\Gamma}(E)$, the differential probability can be presented as follows:
\begin{eqnarray}\label{eq:despair}
\frac{dP(E)}{dE} & \sim & \overline{\Gamma}(E) \left|  \int_{0}^{\delta t} e^{i (E - \epsilon_{2p} - 2 \omega) \tau} d\tau \right|^2 \\ \nonumber
& & \sim \overline{\Gamma}(E) \left| \frac{\sin (E - \epsilon_{2p} - 2 \omega) \delta t}{(E - \epsilon_{2p} - 2 \omega) \delta t} \right|^2.
\end{eqnarray}
While $\overline{\Gamma}(E)$ corresponds only to the energy conservation, the second factor 
approximately describes the interference of the ionized electrons, similar to the 
considerations of Ref.~\citep{arbo2010}. The minimum appearing in the spectra 
corresponds to the first node of this factor: with the increase of the peak 
width $\delta t$ (which can be done by increasing either 
the intensity or the pulse duration) the position of the minimum shifts to the 
main peak in the spectrum according to Eq.~\eqref{eq:despair}, which is clearly 
seen in the results for the small peak intensity 
$I = 10^{10}$ W/cm$^2$ and pulses containing 13 to 20 optical cycles (see 
Fig.~\ref{fig:spectra_l}).
\begin{table}
\caption{The phase shifts $\Delta\Phi_{\mathrm{R}}$ and $\Delta\Phi$ defined by 
Eq.~\eqref{eq:dedtr} and Eq.~\eqref{eq:dedt} respectively, calculated for 
different laser peak intensities.}
\label{table:dedt}
\begin{ruledtabular}
 \begin{tabular}{c c c} 
 Peak intensity, W/cm$^2$ & $\Delta\Phi_{\mathrm{R}} / 2 \pi $ & $\Delta\Phi / 
2 \pi $ 
\\ [0.5ex] 
 \hline
 2.5 $\times 10^{11}$ & 1.08 & 1.09 \\
 3.0 $\times 10^{11}$ & 0.99 & 1.09 \\
 3.5 $\times 10^{11}$ & 0.95 & 1.05 \\
 4.0 $\times 10^{11}$  & 0.91 & 1.04 \\
 4.5 $\times 10^{11}$  & 0.88 & 1.00 \\
 5.0 $\times 10^{11}$  & 0.88 & 1.03 \\
 5.5 $\times 10^{11}$  & 0.86 & 1.02 \\ [1ex] 
\end{tabular}
\end{ruledtabular}
\end{table}

For the other pulse envelope functions, like Gaussian or sine-squared, 
the effects caused by the edges of the envelope may be 
more significant, leading to emergence of complex interference structures 
in the spectra~\cite{telnov1995, wickenhauser2006}. However, as we have checked 
by performing calculations with the sine-squared pulse envelope function, the 
interference mechanism studied here remains dominant for this pulse shape as 
well, at least for relatively long pulses. Insensitivity of the Rabi-flopping 
interference pattern in the photoelectron spectrum to the pulse shape could 
facilitate its observation in the experiments.
\begin{figure}
\includegraphics[width=1\linewidth]{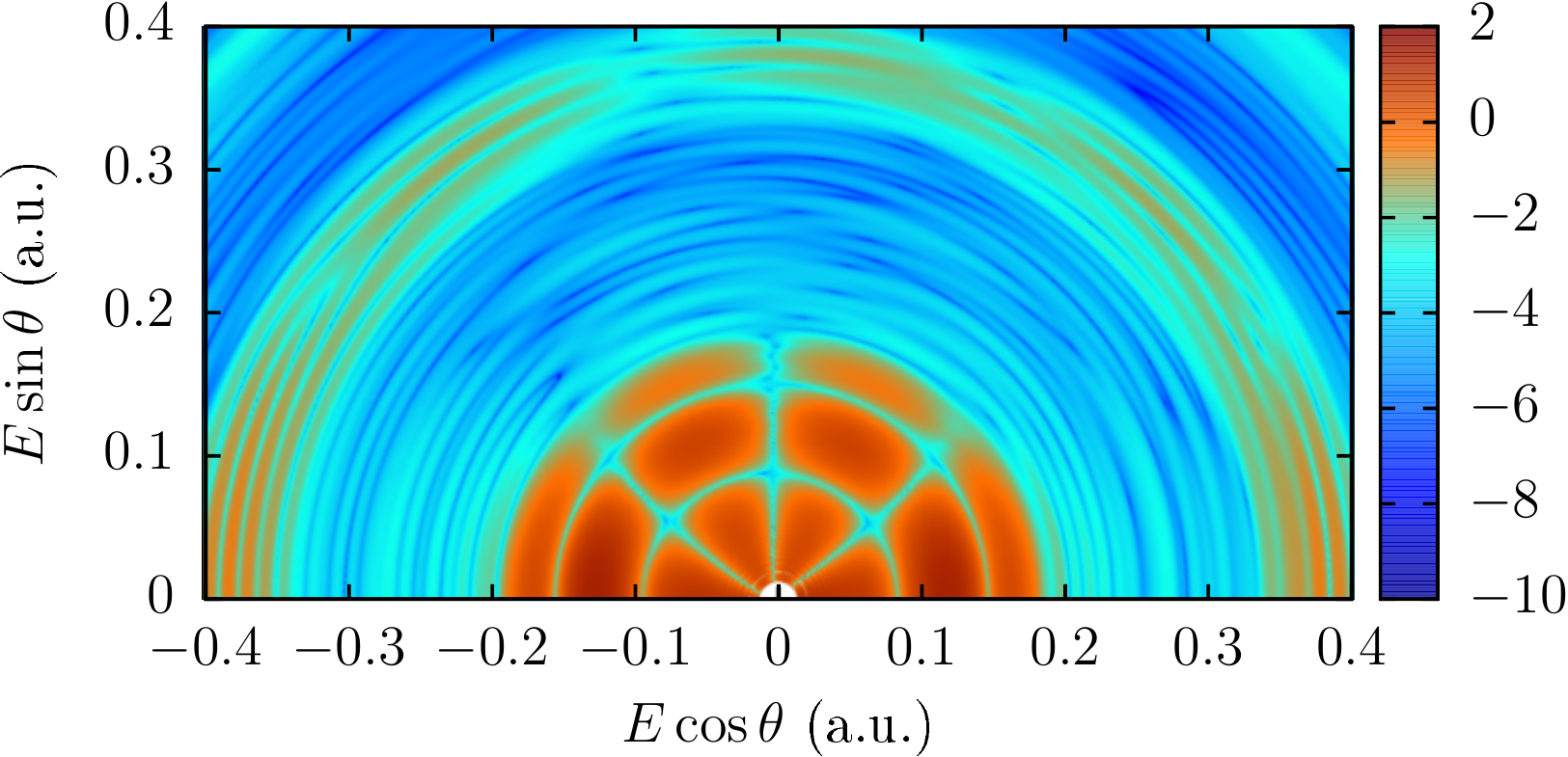}
\caption{Photoelectron angular and energy distribution 
after above-threshold ionization of a Li atom exposed to a linearly-polarized 
laser pulse with peak intensity $I = 5 \times 10^{11}$~W/cm$^2$. The pulse 
contains 20 optical cycles, the laser wavelength is 671~nm and corresponds 
to the resonance between the $2s$ and $2p$ states. The pulse envelope is given 
by Eq.~\eqref{eq:modelenv}. The PAD intensity scale is logarithmic and shown as 
a color map.}
\label{fig:pad}
\end{figure}

In Fig.~\ref{fig:pad}, we present the PAD after multiphoton 
ionization of the Li atom calculated by Eq.~\eqref{eq:pad} for the laser peak 
intensity $I = 5 \times 10^{11}$~W/cm$^2$. The angular distribution has a 
well-known ring structure. Here we show the first two rings, corresponding to 
the ionization by three and four photons. As known from the literature (see, 
e.g., Ref.~\cite{chen2006}), the number of nodes in the angular distribution 
equals to the dominant value of the angular momentum $l$ in the final 
continuum state. For the first ring in the PAD, the dominant $l = 3$, as 
anticipated, since only three photons are required for the ionization. The 
radial structure of the stripes corresponds to the interference mechanism 
discussed above. As one can see, it is independent of the electron emission 
angle and reproduces the same features as the photoelectron energy spectrum.

\section{Conclusion}\label{sec:summary}
In this paper, we have presented photoelectron angular distributions and energy 
spectra after multiphoton above-threshold ionization of Li atoms in the 
one-photon Rabi-flopping regime. The Li atom is described by the 
single-active-electron model with a quality core potential, which reproduces 
accurately the excitation and ionization energies, as well as transition dipole 
matrix elements. The interaction with the linearly-polarized laser field is 
treated in the dipole approximation using the length gauge. The time-dependent 
Schr\"{o}dinger equation is solved efficiently with the help of the 
time-dependent generalized pseudospectral method. The calculations have been 
performed for the laser peak intensities in the range 
1~$\times$~10$^{\mathrm{11}}$ to 5.5~$\times$~10$^{\mathrm{11}}$~W/cm$^2$. The
carrier wavelength is set to 671~nm, so the photon energy matches the 
experimental transition energy between the $2s$ and $2p$ states of the Li 
atom.

We have shown that the population transfer between the 
ground $2s$ and excited $2p$ states in the resonant laser 
field is reflected in the photoelectron energy spectra which manifest 
interference oscillatory structures with the spacing 
between the adjacent minima equal to the Rabi frequency $\Omega$. The main 
ionization channel is controlled by the excited state population and switched 
on at specific moments in time when the ionization rate is the highest, thus
implementing the double-slit interference picture in the time 
domain~\cite{lindner2005}. The transformations of the interference structures 
with the increase or decrease of the pulse area have been also revealed and 
analyzed.

For all our calculations reported in this paper, we used the trapezoidal pulse 
envelope function to minimize the interference effects in 
the electron spectra related to the pulse shape~\cite{telnov1995, 
wickenhauser2006} and not caused by the Rabi flopping. However, we have also 
performed similar calculations for the sine-squared pulse envelope and found 
that the interference pattern due to the resonant population transfer in the 
Rabi-flopping regime is still dominant. We should also note that the 
interference structures emerging in the electron spectra in the Rabi-flopping 
regime are not specific to the Li atom and can be observed for the other atomic 
or molecular targets with similar properties of the electronic energy levels.

\begin{acknowledgments}
This investigation was supported by Saint Petersburg State
University (SPbSU) and  Deutsche Forschungsgemeinschaft (DFG) 
(Grants No. 11.65.41.2017 and No. STO 346/5-1). 
D. A. T. acknowledges the support from the German-Russian Interdisciplinary
Science Center (G-RISC) funded by the German Federal
Foreign Office via the German Academic Exchange 
Service (DAAD), and from TU Dresden (DAAD-Programm Ostpartnerschaften). 
The calculations were performed at the Computing Center of SPbSU Research Park.
\end{acknowledgments}

\end{document}